\documentclass[letter]{aa}  

\usepackage{graphicx}
\usepackage{txfonts}
\usepackage[english]{babel}
\usepackage{amssymb}
\usepackage{amsmath}
\usepackage{mathrsfs}
\usepackage{latexsym}
\usepackage{longtable}
\usepackage{multirow}
\usepackage{epsf}
\usepackage{color}
\usepackage{float}
\usepackage{enumerate}
\usepackage{txfonts}
\usepackage{natbib}
\usepackage{comment}
\usepackage{capt-of}
\usepackage{rotating}
\usepackage{ulem}

\newcommand{\as}{\arcsec{}}

\begin{document} 

   \title{Spirals inside the millimeter cavity of transition disk SR 21
\thanks{Based on observations performed with SPHERE/VLT under program ID 1100.C-0481(Q).}}

\author{G.\,A.\,Muro-Arena\inst{1}
\and C.~Ginski\inst{1,2}
\and C.~Dominik\inst{1}
\and M.~Benisty\inst{3,4}
\and P.~Pinilla\inst{5}
\and A.\,J.\,~Bohn\inst{2}
\and T.~Moldenhauer\inst{6}
%\and D.~Thun\inst{5}
\and W.~Kley\inst{6}
\and D.~Harsono\inst{2}
\and T.~Henning\inst{5}
\and R.\,G.\,~van Holstein\inst{2,7}
\and M.~Janson\inst{8}
\and M.~Keppler\inst{5}
\and F.~M\'enard\inst{3}
\and L.\,M.\,~P\'erez\inst{9}
\and T.~Stolker\inst{10}
\and M.~Tazzari\inst{11}
\and M.~Villenave\inst{3}
\and A.~Zurlo\inst{12}
\and C. Petit\inst{13}
\and F. Rigal\inst{1}
\and O. M\"oller-Nilsson\inst{5}
\and M. Llored\inst{14}
\and T. Moulin\inst{3}
\and P.  Rabou\inst{3}}

\institute{Anton Pannekoek Institute for Astronomy, University of Amsterdam, Science Park 904,1098XH Amsterdam, The Netherlands \email{g.a.muroarena@uva.nl}
\and Leiden Observatory, Leiden University, PO Box 9513, 2300 RA Leiden, The Netherlands
\and Univ. Grenoble Alpes, CNRS, IPAG, 38000 Grenoble, France
\and Unidad Mixta Internacional Franco-Chilena de Astronom\'{i}a (CNRS, UMI 3386), Departamento de Astronom\'{i}a, Universidad de
Chile, Camino El Observatorio 1515, Las Condes, Santiago, Chile
\and Max Planck Institute for Astronomy, K\"onigstuhl 17, 69117, Heidelberg, Germany
\and Institut für Astronomie und Astrophysik, Universit\"at T\"ubingen, Auf der Morgenstelle 10, D-72076 T\"ubingen, Germany
\and European Southern Observatory, Alonso de Córdova 3107, Casilla
19001, Vitacura, Santiago, Chile
\and Department of Astronomy, Stockholm University, Stockholm, Sweden
\and Universidad de Chile, Departamento de Astronomia, Camino El Observatorio 1515, Las Condes, Santiago, Chile
\and Institute for Particle Physics and Astrophysics, ETH Zurich, Wolfgang-Pauli-Strasse 27, 8093 Zurich, Switzerland
\and Institute of Astronomy, University of Cambridge, Madingley Road, CB3 0HA, Cambridge, UK
\and N\'ucleo de Astronom\'ia, Facultad de Ingenier\'ia y Ciencias, Universidad Diego Portales, Av. Ejercito 441, Santiago, Chile
\and DOTA, ONERA, Université Paris Saclay, F-91123, Palaiseau France 
\and Aix Marseille Universit\'e, CNRS, CNES,  LAM, Marseille, France
}

\date{}

  \abstract
   {Hydrodynamical simulations of planet-disk interactions  suggest that planets may be responsible for a number of the substructures frequently observed in disks in both scattered light and dust thermal emission. Despite the ubiquity of these features, direct evidence of planets embedded in disks and of the specific interaction features like spiral arms within planetary gaps still remain rare.}% \mb{if it's rare, it's not absent, pick one of the two. i would say 'spiral arms within planetary gaps'.}}
   {In this study we discuss recent observational results in the context of hydrodynamical simulations in order to infer the properties of a putative embedded planet in the cavity of a transition disk.}
   {We imaged the transition disk SR 21 in H-band in scattered light with SPHERE/IRDIS and in  thermal dust emission with ALMA band 3 (3\,mm) observations at a spatial resolution of 0.1\arcsec{}. We combine these datasets with existing band 9 (430\,$\mu$m) and band 7 (870\,$\mu$m) ALMA continuum data. 
   %and compare them to previous results from hydro simulations to explain the most prominent features seen in this disk at different wavelengths.
   }
   {The Band 3 continuum data reveals a large cavity and a bright ring peaking at 53\,au strongly suggestive of dust trapping. The ring shows a pronounced azimuthal asymmetry, with a bright region in the north-west that we interpret as a dust over-density. A similarly-asymmetric ring is revealed at the same location in polarized scattered light, in addition to a set of bright spirals inside the mm cavity and a fainter spiral bridging the gap to the outer ring. These features are consistent with a number of previous hydrodynamical models of planet-disk interactions, and suggest the presence of a $\sim$1\,M$_{\rm{Jup}}$ planet at 44\,au and PA=11deg.  
   This makes SR21 the first disk showing 
   %\paola{a hint of} 
   spiral arms inside the mm cavity, as well as one for which the location of a putative planet can be precisely inferred.} %\mb{I would like to remove the last sentence.} \paola{Yes, I agree with Myriam that the last sentence should be removed.}}
   {SR\,21's main features in both scattered light and thermal emission are consistent with hydrodynamical predictions of planet-disk interactions. With the location of a possible planet being well-constrained by observations, it is an ideal candidate for follow-up observations to search for direct evidence of a planetary companion still embedded in its disk.}

\titlerunning{Spirals inside the cavity of transition disk SR21}
\authorrunning{Muro-Arena et al.}
\maketitle

\section{Introduction}
\label{sec:introduction}

Protoplanetary disks are the places where planet formation takes place, yet the associated timescales and evolutionary processes are not well-constrained by observations. Planet formation is thought to induce morphological features in their parent disk, such as rings and gaps, spiral arms or vortices.
Such structures were indeed detected in a growing sample of disks in mm-emission as well as in optical and near infrared scattered light (see e.g. \citealt{2018ApJ...869L..41A, 2018ApJ...863...44A} for recent examples). If indeed caused by embedded planets, this suggests that planet formation occurs early in the lifetime of the disk. Of particular interest in that respect have been so-called transition disks. These are disks that show a lack of flux in the 10\,$\mu$m wavelength range, suggestive of large cavities. 

It has been proposed that these cavities are carved out by the emerging proto-planets, yet so far only one such system is known where planets are indeed detected (PDS\,70, \citealt{2018A&A...617A..44K, 2018A&A...617L...2M, 2019NatAs...3..749H}). Thus additional observations are required to make firm the link between planet formation and the opening of cavities.

EM*\,SR\,21 (hereafter, SR\,21) is a nearby (138.4$\pm$1.1\,pc, \citealt{2018yCat.1345....0G}) member of the Ophiuchus star forming region. 
SR\,21 has a wide (6.7\arcsec{}) binary companion \citep{2003ApJ...591.1064B} confirmed by recent GAIA DR2 measurements of proper motion and parallax. The presence of a close ($\sim$0.1\arcsec{}) companion (possibly a forming proto-planet) was suggest by \cite{2009ApJ...698L.169E} using mid-IR aperture masking interferometry data, but was so far not confirmed.

A low resolution spectroscopic study by \cite{2014ApJ...786...97H} found SR\,21 to be of spectral type F7 with an age of 10\,Myr and a mass of 1.67\,M$_\odot$. Higher resolution X-SHOOTER observations by \cite{2015A&A...579A..66M} confirmed its status as an intermediate mass star.

The SR\,21 system has been the subject of intense observational study. \cite{2007ApJ...664L.107B} utilized Spitzer spectroscopic observations in the IR to classify it as a transition disk with a large inner cavity. \cite{2011ApJ...732...42A} indeed found an inner cavity radius of the mm-sized grains of $\sim$40\,au using SMA observations at 880\,$\mu$m. This is confirmed with higher spatial resolution ALMA observations at 430\,$\mu$m by \cite{2014ApJ...783L..13P}, who also reports a large asymmetry in the resolved mm-emission, suggestive of a vortex and a potential spiral structure in the South-West of the disk. It was found by \cite{2008ApJ...684.1323P} that the gas extends much further inwards to a truncation radius of $\sim$7\,au.

The system was previously observed at near infrared wavelengths by \cite{2013ApJ...767...10F} who detect morphological differences in polarized scattered light. They found that the small dust grain population extends inside the mm-cavity. While their data was suggestive of a potential warp in the disk due to a possible change in disk position angle with radius, they did not resolve specific structures inside the mm-cavity. A very recent study by \cite{2019arXiv190807427S} used non redundant masking imaging data in the near infrared to find a truncation radius of small dust grains between 4\,au and 7\,au, consistent with the findings of \citet{2008ApJ...684.1323P}. Their data is also suggestive of strong asymmetries in the inner disk, caused either by a warp or spiral features.

We present here new polarized scattered light observations of the SR\,21 system carried out with VLT/SPHERE (Spectro-Polarimetric High-contrast Exoplanet REsearch, \citealt{2019A&A...631A.155B}) in the near infrared and with ALMA in Band 3 at 3\,mm.

\section{Observations and data reduction}
\label{sec:datared}

In the following we describe the details of the observations of the SPHERE and ALMA observations as well as the data reduction.

%[Details on the observations and the data reduction for bothof these datasets are described in section 2. Section 3 outlines theobservational results, which are then discussed in the context ofhydrodynamical simulations of planet-disk interaction in section4. Our conclusions are summarized in section 5.]

\subsection{Observations}

The SPHERE observations were acquired on March 1st 2018 with SPHERE/IRDIS (Infra-Red Dual-beam Imager and Spectrograph, \citealt{2008SPIE.7014E..3LD}) in dual-beam polarimetric
imaging mode (dpi, \citealt{2014SPIE.9147E..1RL, vanHolstein2020, deBoer2020}).
We observed the system with the broad band H filter and the broad band J filter. H-band observations were carried out with a Lyot coronagraph in place with an inner working angle of 92.5\,mas (\citealt{2009A&A...495..363M, 2011ExA....30...39C}).
The conditions during both observations sequences were excellent with Seeing varying between 0.44\arcsec{} and 0.57\arcsec{} and coherence time above 4\,ms. For H-band we recorded 21.3\,min of data while for J-band we recorded 13.3\,min.

%The conditions during these observations were excellent with an average Seeing of 0.44\,arcsec and a coherence time of the atmosphere of 5.36\,ms. We recorded 5 polarimetric cycles, each consisting of one Q$^+$, Q$^-$, U$^+$ and U$^-$ exposure, each with an exposure time of 64\,s yielding a total integration time of 21.3\,min.
%J-band observations were carried out without coronagraph and with a neutral density filter in place to prevent saturation. Conditions remained good with an average Seeing of 0.57\,arcsec and a coherence time of 4.21\,ms. We again recorded 5 polarimetric cycles. For each sub-set of the polarimetric cycle we recorded 20 exposures with an individual exposure time of 2\,s.

The ALMA Band 3 observations were carried out in Cycle 5 (2017.1.00884.S, PI: Pinilla) on November 9th 2017 with the 12\,m array in configuration C43-8 using 44 antennas. The total integration time was 41\,min with a maximum baseline of 8.5km. We note that only these long baseline observations were carried out and we are missing shorter baselines, limiting the maximum recoverable scale to 1.4\,arcsec. The spectral setup had two spectral windows centered on the $^{13}$CO\,\textit{J}=1-0 and C$^{18}$O\,\textit{J}=1-0 transitions with rest frequencies at 110.201354 and 109.782176 GHz, and a bandwidth of 937.5 MHz (resolution of 488 kHz) and 234.38 MHz (resolution of 122 kHz), respectively, and two spectral  windows for the continuum, centered at 108 GHz. In one of them, the correlator was set to time division mode (128 channels, 31.25 MHz resolution and 1875 MHz total bandwidth), while in the other one it is set to frequency division mode (3840 channels, 488 kHz resolution, 937.5 MHz total bandwidth). 

%\paola{I saw too late that you have a sub-section in data reduction, you can move part of the previous paragraph to the next section. }

\subsection{Data reduction}
\begin{figure*}[ht]
\center
%\begin{tabular}{ccc}
\includegraphics[width=0.98\textwidth]{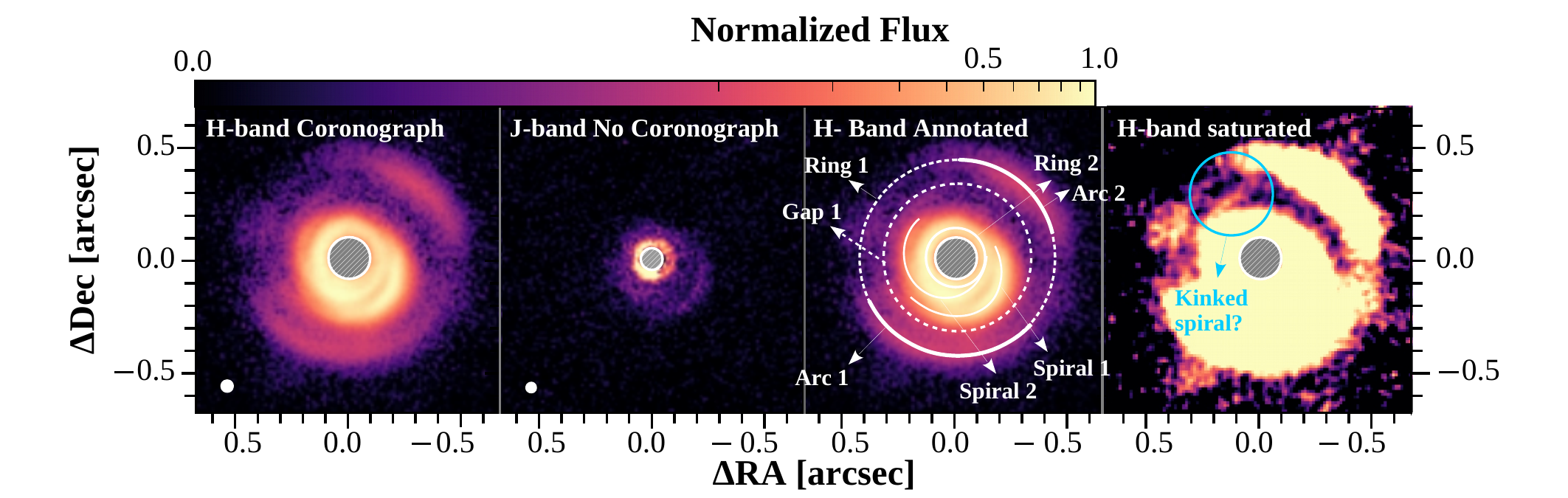} 
\caption{SPHERE/VLT scattered light observations of SR\,21. From left to right we show the H-band observation with coronagraph, the J-band observation without coronagraph, the annotated H-band data, each normalized to its maximum flux. The rightmost panel shows an over-saturated cut of the $r^{2}$-scaled H-band data. We are showing in all cases the Q$_\phi$ image. In the H-band epoch the grey hatched software mask indicates the coronagraph, while in J-band it indicates the inner working angle of the polarimetric observations.
In both epochs we indicate the size of one resolution element with white filled circles in the lower left corner.} 
\label{fig:SPHERE}
\end{figure*}

The SPHERE observations were reduced with the IRDAP (IRDIS Data reduction for Accurate Polarimetry) pipeline by \citet{vanHolstein2020}. The pipeline follows largely the polarimetric data reduction as outlined in \cite{2016A&A...595A.112G} and \cite{ 2017SPIE10400E..15V}. It includes a full Mueller Matrix model of the instrument and telescope system and thus can accurately compute and subtract the instrumental polarization. After the initial data reduction to retrieve Stokes Q and U images, the pipeline computes the polar Stokes vector images Q$_\phi$ and U$_\phi$ following \cite{2006A&A...452..657S}. %Q$_\phi$ contains all azimuthally aligned polarimetric signal as positive values and radially aligned polarization signal as negative values. U$_\phi$ contains all polarization signal 45$^\circ$ offset from azimuthal or radial orientation. For small inclinations (as is the case in the EM*\,SR\,21 system) it is expected that the signal is dominated by single scattering and thus (close to) all polarimetric signal should show azimuthal alignment and is thus contained in the Q$_\phi$ image.
We show the final Q$_\phi$ images for the SPHERE H and J-band observations in figure~\ref{fig:SPHERE}. 
%In addition, we created a combined image of both epochs by multiplying the J-band observation with the inverse coronagraph profile (Wilby et al., submitted) and adding it after appropriate scaling for exposure times to the H-band observation. This combined image is also shown in figure~\ref{fig:SPHERE}.\\

Self-calibration was performed on band 3 using \texttt{CASA v. 5.1.1} \citep{2007ASPC..376..127M}, improving the signal-to-noise of the data by a factor of two compared to the product. We used \texttt{tclean} to produce the images of Band 3 data, with Briggs weighting and a robust parameter of 0.0 for the best compromise between resolution and sensitivity. The synthesized beam of the image is 0.10\arcsec{}$\times$0.09\arcsec{} with an rms of 13.6\,$\mu$Jy/beam, a total flux of 8.2\,mJy, with a peak value of 0.35\,mJy/beam ($\sim25\times \sigma$). Previous observations at 3.3\,mm using ATCA (Ricci et al., 2010) found a total flux of 4.2\,mJy, our current observations at similar wavelength suggest twice the ATCA flux. We also use the data presented in Pinilla et al. 2015 in Band 9 (690 GHz) and Band 7 (343 GHz) from Cycle 0 and Cycle 1 ALMA observations 
\citep{2014ApJ...783L..13P, 2016A&A...585A..58V}.

%\mb{To be updated (self cal?): The ALMA observations were  calibrated through the  calibration pipeline in the Common Astronomy Software Applications (CASA 5.1.1; McMullin et al. 2007). The images were obtained using the \texttt{tclean} task, and imaged using Briggs weighting with a Robust parameter of XX. The synthetized beam size achieved
%are 0.109\arcsec{}$\times$0.094\arcsec{}, with a PA of 89.84 $^\circ$. The achieved sensitivity in Band 3 is XY µJy beam$^{−1}$.}

\section{Disk morphology} 
\label{sec:images}

In the following sections we describe the morphology of the disk as seen in scattered light and (sub-)mm emission and compare it with hydrodynamic models of planet-disk interaction.

\subsection{Scattered light images}

%3xgaussian fit:
%Ring:
%peak: 0.4\as, width: 0.078\as (sigma)  0.18\as %(FWHM)

%1xgaussian fit:
%Ring:
%peak: 0.4\as, width: 0.068\as (sigma) 0.16\as (FWHM)

%Disk outer radius:

%1-sigma: 0.585\as
%2-sigma: 0.533\as 
%3-sigma: 0.51\as

The H-band polarized scattered light images of SR21 (see Fig.\,\ref{fig:SPHERE}, left panel) reveal the complex structure of a nearly face-on disk. An asymmetrical outer ring (Ring\,1, third panel Fig.\,\ref{fig:SPHERE}) is detected at a radius of $\sim$0.4\as, in a similar location to the continuum emission previously detected in Band\,7 \citep{2016A&A...585A..58V} and Band\,9 ALMA data \citep{2014ApJ...783L..13P}. This ring is constituted of two bright arcs, along the north-west quadrant (Arc\,2) and the south side of the disk (Arc\,1), also seemingly coincidental with the two brighter arcs detected in Band\,7 by \citet{2016A&A...585A..58V}. We derive an outer radius of 0.51\as ($\sim$70\,au), at the 3-$\sigma$ level, from the (azimuthally-averaged) radial polarized intensity profile. 

In more detail, the H-band data reveals two bright spiral arms that are visible inside this ring, winding out clockwise %(West-from-North), 
and with the outer one (Spiral\,1) being the brightest of the two. They extend from an inner radius of $\sim$0.14\as (Ring\,2), inside the mm-cavity previously detected with both SMA \citep{2011ApJ...732...42A} and ALMA in multiple bands, and rapidly become fainter with increasing radius. A partial gap (Gap\,1) or shadow separates them from the outer ring. 

%\mb{it's fine but spiral 1 doesn't have the typical shape of a spiral, it's clear in the polar plot. a typical spiral doesn't come back with radius but we might have projection effects and also another feature.}\gaby{we mention this in a few paragraphs}

The second panel of Fig.\,\ref{fig:SPHERE} shows the J-band non-coronagraphic epoch, where we can see the polarized scattered light extending in as far as $\sim$0.06\as ($\sim$8\,au) from the star. Here we see a third bright ring at this location, consistent with the presence of a gas and micron dust grain cavity in the innermost few au as determined by \citet{2008ApJ...684.1323P}.

The right panel of Fig.\,\ref{fig:SPHERE} shows one faint additional feature in an over-saturated cut of the H-band data, in the North side and inside the gap (Gap\,1) between the spiral arms and Ring\,1. What appears to be a kinked spiral, or possibly a streamer, can be seen wrapping out clockwise, extending from one of the bright inner spirals in the north-east, and out towards the Arc\,2 in the north-west. Using the U$_\phi$ image (which contains little or no astrophysical signal) to generate a radial noise map, we find that this feature has a signal-to-noise ratio of $\sim$3.5 close to the launching point at the inner ring. This signal-to-noise ratio decreases to $\sim$2.3 near the point where it connects to the outer ring.
We note that this signal-to-noise ratio was computed on a pixel by pixel basis and takes not into account that a clear shape is visible in the data.

\begin{figure}[ht]
\center
%\begin{tabular}{ccc}
\includegraphics[width=87mm]{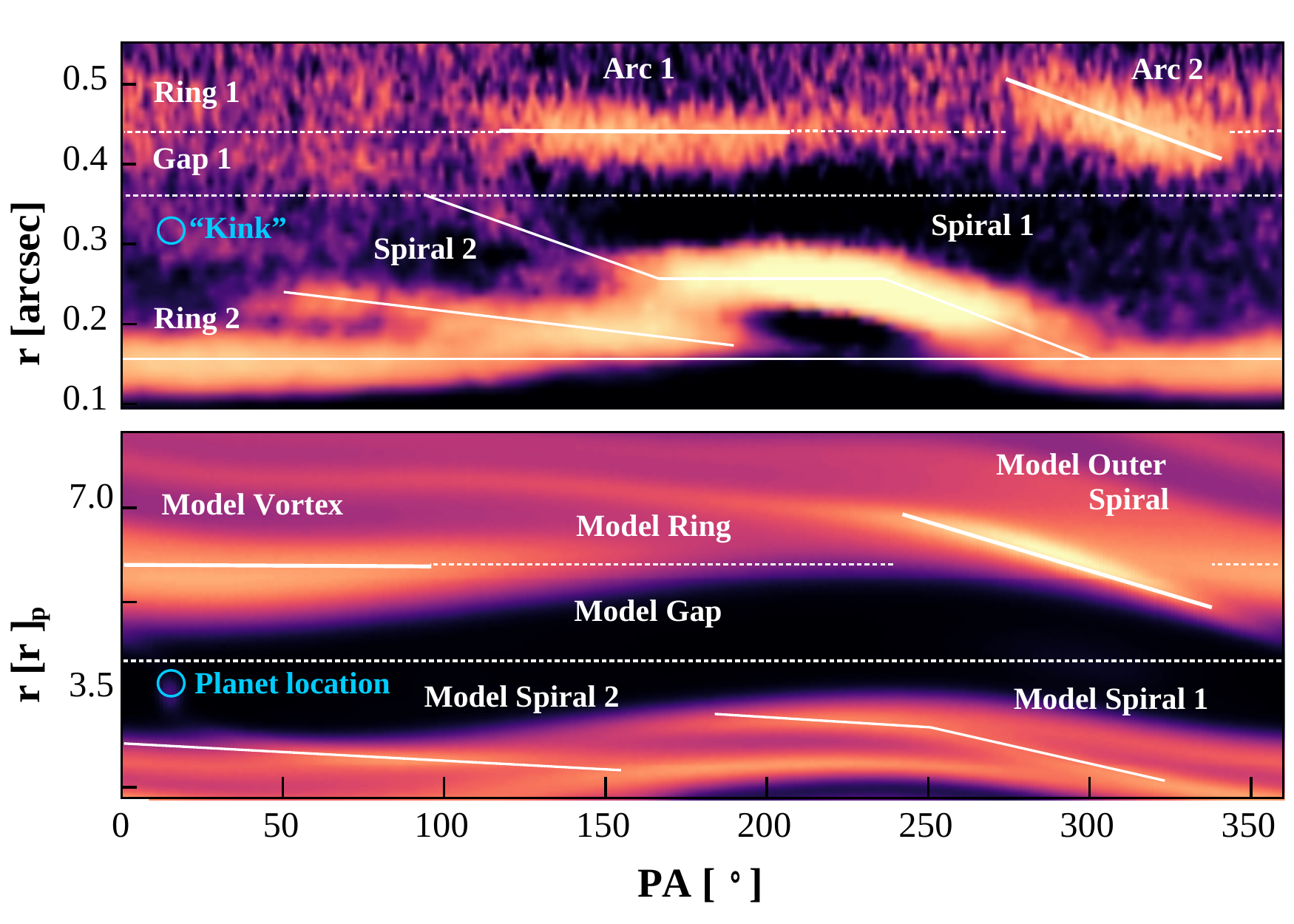} 
\caption{Projection of the SPHERE scattered light image into a polar coordinate system (top panel). Position angle is measured counter-clockwise from the North direction. The SPHERE data was scaled by the square of the separation in order to compensate for the drop in flux due to illumination effects. We then high-pass filter prior to re-projection in order to highlight the disk features. Bottom panel shows the polar projection \textbf{of the gas surface density} of one of the hydrodynamical models from \cite{pinilla2015}. The model assumes $\alpha_{\rm{turb}}$= 10$^{-3}$, planet mass ratios of 10$^{-4}$ and 10$^{-3}$ with respect to the central star and a ratio of orbital radii of 3.5. We mark the location of the outer planet with a circle in the model image and the location of the kinked spiral in the data in the top panel.
%add ref properly
} 
\label{fig:sphere-polar}
\end{figure}

The polar projection of the high-pass filtered H-band image can be seen in the top panel of Fig.\,\ref{fig:sphere-polar}. The projection showcases the spiral arms, seen here as two diagonal bright lanes, with Spiral\,1 showing a varying pitch angle. In this figure we can also see that the Arc\,2 region of Ring\,1, around 300 to 350$^{\circ}$ in PA, is not entirely horizontal. Ring\,1 is therefore not entirely circular, with the north-western arc appearing like a spiral feature in the polar projection. A similar behavior can be seen, for example, in the hydrodynamical models from \cite{pinilla2015}, one of which is shown in the bottom panel of Fig.\,\ref{fig:sphere-polar}. This model shows the resulting gas surface density from the interaction of two planets in the disk with $\alpha_{\mathrm{turb}}$= 10$^{-3}$, with mass ratios of 10$^{-4}$ and 10$^{-3}$ for the planets located at 1\,$r_{\mathrm{p}}$ and 3.5\,$r_{\mathrm{p}}$, respectively. The location of the outer planet in the model is marked with a circle in the bottom panel, and is plotted in alignment with the PA location of the kink in the faint spiral feature in the H-band data (marked as a circle in the middle panel). 
Several features of the hydrodynamical model correspond well with the observations. Going from small to large radii we see that the azimuthal location and pitch angle of spiral 1 and 2 in the observation match well with the model inner spiral 1 and 2 marked in the bottom panel. The eccentric gap seen in the model seems at least in part consistent with the observations between PA of $\sim$200$^{\circ}$ and 360$^{\circ}$. At smaller position angles the observational data is too noisy to confirm the shape of the gap. The observational feature in the outer scattered light ring marked as Arc\,1 in the figure may well correspond to the vortex seen in the model. The azimuthal vortex location in the model is not fixed relative to planet position as can be seen in \cite{pinilla2015}. The azimuthal position and pitch angle of Arc\,2 in the observation match exceptionally well with the corresponding model outer spiral.

%\begin{figure}[!h]
%\center
%\begin{tabular}{ccc}
%\includegraphics[width=87mm]{SR21_saturated_v4.pdf} 
%\caption{SPHERE Q$_\phi$ image with an r$^2$-scaling, binned by a factor of 2 and shown on a strongly saturated color map. A faint, possibly kinked, spiral arm or streamer is visible in the North.} 
%\label{fig:sphere-saturated}
%\end{figure}

\subsection{Band 3 dust continuum}

\begin{figure*}[ht]
\center
%\begin{tabular}{ccc}
\includegraphics[width=0.98\textwidth]{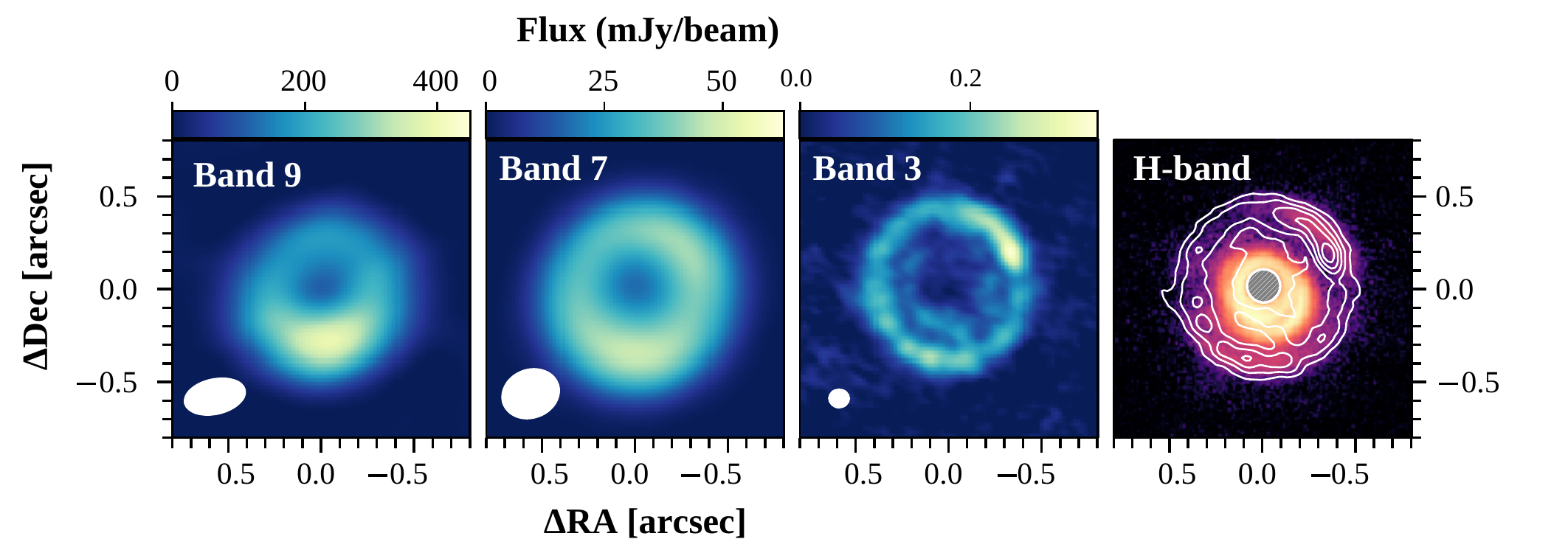} 
\caption{ALMA dust continuum images of SR\,21. Band\,9 \textbf{(rms= 3\,mJy/beam)} and Band\,7 \textbf{(rms= 0.2\,mJy/beam)} were reproduced from \cite{2014ApJ...783L..13P} and \cite{2015A&A...584A..16P}, respectively. We add our new Band\,3 observations \textbf{(rms= 13.6\,$\mu$Jy/beam)} in the third panel. Beam size and orientation are indicated by the white ellipses in each panel. Right panel shows the H-band data \textbf{in the same color scale as Fig\,\ref{fig:SPHERE}, left panel}, with the ALMA Band\,3 contours at 6, 12, 18 and 24\,$\sigma$ overlaid.} 
\label{fig:ALMA}
\end{figure*}
%1-sigma outer radius 0.579\as
%2-sigma outer radius 0.5495\as
%3-sigma outer radius 0.5325\as

%fit from 0.33\as to 0.66\as
%gaussian fit: peak: 0.387\as, width: 0.08\as (sigma) 0.1884\as (FWHM)
%2xgaussian fit: peak: 0.4\as, width: 0.066\as (sigma), 0.155 \as (FWHM)

The Band\,3 continuum map of SR\,21 is shown in the third panel of Fig.\,\ref{fig:ALMA}. The image shows a narrow ($\sim$0.18\as FWHM) ring centered at 0.387\as (determined from a Gaussian fit to the azimuthally-averaged radial profile between 0.33\as and 0.66\as). This ring largely overlaps with the Ring\,1 feature seen in scattered light (Fig.\,\ref{fig:ALMA}), and is brightest in the north-west, at the location of Arc\,2. A fragmented or discontinuous second ring, previously unresolved in Band\,9 and Band\,7 observations (Fig.\,\ref{fig:ALMA}), is seen inside it at a radius of $\sim$0.23\as. This inner ring partially overlaps with the Spiral\,1 feature seen in scattered light, as seen in the left panel of Fig.\,\ref{fig:ALMA}. 

%\mb{Over which radii, maybe that's why Spiral 1 is so weird because it might be a combination of a ring and a spiral.}

%\begin{figure}[!h]
%\center
%\begin{tabular}{ccc}
%\includegraphics[width=87mm]{fig2.pdf} 
%\caption{Top: Band 3 ALMA continuum, normalized to peak intensity. Beam size is indicated in the left bottom corner. Bottom: SPHERE H-band $Q_{\phi}$ data, with the band 3 ALMA continuum contours overlaid. Contours indicated with solid lines, at levels of 5, 10, 15, 20, 25 and 30-$\sigma$.} 
%\label{fig:ALMA}
%\end{figure}

%\begin{figure}[!h]
%\center
%\begin{tabular}{ccc}
%\includegraphics[width=87mm]{fig3.pdf} 
%\caption{Azimuthally-averaged radial profiles for SPHERE H-band (black) and Band 3 ALMA (red). Vertical dashed lines indicate the location of Ring 1 in each band. The error bars correspond to the standard deviation measured from the $U_{\phi}$ data (H-band) and the continuum rms (band 3). The horizontal lines in the labels indicate the resolution of each dataset.} 
%\label{fig:profiles}
%\end{figure}

\section{Discussion}
\label{sec:discussion}

In the following sections we discuss the observed morphology of the disk in the context of similar observations. We in particular discuss the hypothesis that one or multiple planets are responsible for the observed disk features.

\subsection{Comparison to similar objects}

Spirals in disks are meanwhile known to be a common feature in scattered light observations. Typically in cases where spirals are observed these are the dominating structures in the disk, see e.g. the HD\,135344\,B system (\citealt{2012ApJ...748L..22M,2013A&A...560A.105G,2016A&A...595A.113S}), the MWC\,758 system (\citealt{2013ApJ...762...48G,2015A&A...578L...6B}) or the LkHa\,330 system (\citealt{2016AJ....152..222A,2018AJ....156...63U}). 
In all of these cases the scattered light spirals are launched from an inner ring and are the outermost structures visible. Similarly all of these disks show large opening angles for the spiral structures, varying between $\sim$10 and 16$^{\circ}$ for all three disks \citep{2012ApJ...748L..22M, 2013ApJ...762...48G, 2015A&A...578L...6B, 2015ApJ...809L...5D, 2018AJ....156...63U}. \\
The scattered light image of SR\,21 is remarkably different. While the spirals are still launched from an inner ring-like structure, they are tightly wound, varying between $\sim$14 and 2$^{\circ}$ for Spiral\,1 (pitch angle decreasing with increasing radius) and $<$11$^{\circ}$ for Spiral\,2 (reaching a minimum of close to 0$^{\circ}$ at a radius of $\sim$0.2\as).
%\mb{no absolute need to give numbers here, but did you compare quantitatively, this is an important statement}. 
They are furthermore located inside yet another ring-like structure seen at larger separations. This outer structure in scattered light corresponds to a narrow ring in ALMA Band\,3 observations, the inside of which is mostly devoid of emission and thus largely depleted of mm-sized dust grains. Scattered light spirals inside of ALMA emission are only known for a few disks: HD\,100453 (\citealt{2019MNRAS.tmp.2689R}), V\,1247\,Ori (\citealt{2017ApJ...848L..11K}), HD\,135344\,B (\citealt{2018A&A...619A.161C}) and HD\,169142 (\citealt{2019A&A...623A.140G}). 
%\mb{I think also MWC758 Dong2018, but these are projection effect I think}. 
However, both V\,1247\,Ori and HD\,135344\,B show a very asymmetric, possibly vortex-like, structure which directly connects with the inner scattered light spirals. In both cases, as well as in HD\,169142, there is also further-in mm-emission, so not a cavity as in the case of SR\,21. In the case of HD\,100453 the scattered light structures have corresponding mm-emission features, located at slightly larger radii (\citealt{2019MNRAS.tmp.2689R}), while in HD\,169142 the scattered light spirals are extremely faint and superposed on much brighter ring structures, making them not the main morphological feature. However, in none of these cases, with the exception of HD\,169142, is there a scattered light counterpart detected to the outer mm-emission. 
The detection of the outer ring-like structure in SR\,21 in scattered light also points to an overall low surface height of the inner structures or a strong flaring of the disk, since otherwise the complete outer ring would be shadowed.\\
The correspondence of features in small dust particles and gas surface density has been shown by several studies, e.g. \cite{2015ApJ...809L...5D} and \cite{2016MNRAS.459.2790R}. This is due to the fact that small ($\mu$m-sized) dust particles couple well to the gas, i.e. they generally have a low Stokes number. Large (mm-sized) dust particles seen in emission by ALMA are typically featuring higher Stokes numbers and are thus decoupled from the gas.
The detection of both (inner and outer) structures in scattered light gives thus the rare opportunity for a close comparison of the SR\,21 system with hydrodynamic models of the gas in the disk.

%\paola{here would be nice to point to the hydro model that you found a good match.}. 

\subsection{Indications of planetary companions}

The various features seen in the SPHERE and ALMA images can be explained by one or more planetary companions in the disk. \cite{2018ApJ...869L..42H} and \cite{2018ApJ...869L..46D} show that the rings frequently seen in continuum emission from (sub-)mm grains are likely caused by dust trapping, perhaps due to embedded planets carving gaps in disks, as opposed to being related to snow lines. In the case of SR\,21 the possibility of a low-mass companion being responsible for carving out this cavity is further supported by the small grains seen in scattered light. These are expected to be well-coupled to the gas, tracing spiral arms confined to the mm cavity, as is frequently seen in hydrodynamical models. 

\textbf{Azimuthal asymmetries in the outer ring:} It is hard to tell from scattered light alone whether the asymmetry of the outer ring is caused by azimuthally-localized over-densities in the gas and dust or if they are the product of shadowing by the inner spiral arms. 
%\mb{would only work for Spiral2 right?} 
This asymmetry looks different in all three ALMA bands available: Band\,9 shows an asymmetry in the south, while Bands\,7 and 3 show a ring-like structure with a bright asymmetry at the north. \citet{2015A&A...584A..16P} have already suggested that the southern bright spot seen in Band\,9 may be the result of a hotter region in the southern side of the disk. This is supported by band\,9 being optically-thick (peak optical depth of $\tau\sim$2.0, assuming a physical temperature of 20K). Therefore, continuum emission in this band is not only sensitive to density variations, but also to temperature variations, which can produce structure in the emission. In Band\,3, where the disk is optically thin (peak optical depth of $\tau$= 0.17 assuming a physical temperature of 20K), the asymmetry is reversed, with the southern bright spot disappearing and the northern one becoming prominent. This asymmetry, at this low optical depth, suggests that the northern bright spot seen in SPHERE and ALMA bands\,7 and 3 probably corresponds to a dust over-density at this location of the ring. Band\,7, with an intermediate estimated peak optical depth of $\tau$= 0.5, shows both the over-density and the hot region in the south. Such an overdensity could, for example, be produced in the apastron location of a slightly eccentric disk, where the lower velocity at apastron leads to higher gas densities that can also slow down drifting grains and replicate the overdensity in large grains \citep{2013A&A...553L...3A}.

\textbf{The spiral arms pointing to the likely position of the planet:} The 'kinked spiral' feature we see to the north in the saturated cut of the H-band data (Fig.~\ref{fig:SPHERE}) is strongly-reminiscent of the spirals arising in hydrodynamical models of planet-disk interaction (see e.g. \citealt{2000MNRAS.318...18N}, \citealt{2007A&A...472..981S} or \citealt{2015A&A...584A..16P}). The position of the kink in the spiral, indicating the exact location of the planet in these hydrodynamical models, is found inside of the ALMA ring and in the gap between this ring and the two brighter inner spirals seen with SPHERE (\ref{fig:SPHERE}, right panel). This is consistent with a planet at this location being responsible for carving this gap and thus trapping the larger dust grains in the pressure bump formed outside its orbit. If this is the case, this makes SR21 a prime candidate for the search of a forming planet still interacting with its parent disk, with the location of this planet known a priori and very accurately from the location of the kink in this spiral feature (44\,au, PA ~ 11$^\circ$, see also Figure~\ref{fig:sphere-polar}). 

\textbf{Interaction between spiral and ring:} The kinked spiral feature also appears to lead to the northern bright spot (Arc\,2) along the SPHERE outer ring (Ring\,1), which, as seen in the polar projection (Fig.~\ref{fig:sphere-polar}), is not entirely circular. The figure shows the non-zero pitch angle of Arc\,2, which is very similar to the pitch angle seen in gas density in hydrodynamical simulations of planet-disk interactions where the spiral wave from the planet meets the outer disk. This is seen, for example, in \citet{2015A&A...584A..16P}), in models that were motivated by the Band\,9 ALMA data on SR\,21, and which show the results of the gaseous component of the disk interacting with two planets embedded in the disk. The superposition of the spirals in the gas density between the two planets show a behavior very similar to the bright spirals seen in SR21, with the pitch angle appearing to decrease before increasing again outwards with increasing radius. This complex behavior of the spiral structure might be an indication of a second planet or a binary companion being present in the disk, perhaps in the 7\,au cavity inferred by \cite{2008ApJ...684.1323P} and possibly associated with the signal detected by \cite{2009ApJ...698L.169E}. 

However, we want to point out that there are also similar models available created with just a single planet (see e.g. \citealt{2000MNRAS.318...18N}). From our current data we can thus not conclude on the presence of additional companions inside the scattered light gap.

\subsection{Mass constraints}

The location of Ring\,1 in scattered light vs mm appears to be the same, within uncertainties (the difference in the location of the ALMA and SPHERE rings is $\sim$13\,mas, i.e. a factor 3-4 smaller than the resolution of our H-band data). If trapped in a pressure bump, the location of the large grains is expected to be exterior to that of the small grains observed in scattered light. This radial offset between the rings is larger for larger planetary masses, but can be close to zero for smaller companions, as shown by \cite{2013A&A...560A.111D}. If caused by a planet, its mass must be small and close to 1\,M$_{\mathrm{Jup}}$. Such a small mass planet can still produce the inner spiral arms seen in scattered light, as shown by \cite{2015ApJ...809L...5D}.\\
A small planet mass is also consistent with archival observational data obtained with the Keck telescope and the NIRC2 instrument in the L-band. We discuss this data set in detail in appendix~\ref{keck-appendix}. The Keck observation rules out companions more massive than $\sim$13\,M$_{\mathrm{Jup}}$ at the location of the scattered light gap in SR21.

\section{Conclusions}
\label{sec:conclusions}

EM*\,SR\,21 is a disk exhibiting a number of features consistent with hydrodynamical models of planet formation, including: 
\begin{itemize}
    \item a large cavity in (sub-)mm sized grains, with evidence of dust-trapping at a radius of $\sim$54\,au
    \item two bright spiral arm structures in scattered light located \textit{inside} of this cavity
    \item a scattered light counterpart to the mm ring at $\sim$55\,au
    \item a faint  kinked spiral structure in scattered light, in the gap between the bright inner spirals and the outer ring 
    \item evidence of an over-density in gas and dust in the north side of the disk, where the kinked spiral connects to the outer ring.
\end{itemize} 

Furthermore, the radial location of the gap, outer ring and, in particular, the kink in the faint spiral, allow us to constrain the location of the suspected planet to a radius of $\sim$44\,au and PA of $\sim$11$^{\circ}$. If such a planet is inducing a small eccentricity in the outer disk, it could be causing an over-density of gas in the North side of the $\sim$54\,au ring, as predicted by \citet{pinilla2015}. This over-density could explain the northern bright region of the ring seen in scattered light, as well as the over-density in the larger grains in this region inferred from Bands\,7 and 3 continuum emission. Given the radial location of the ring in H-band and Band\,3, at $\sim$1.2 planetary radii in both datasets, we can rule out a planet of the order of $>$5\,M$_\mathrm{Jup}$, %\paola{I reduced this limit here according to the results from de Juan Ovelar}%
with a lower mass of $\sim$1\,M$_\mathrm{Jup}$ being favored. 

Finally, the irregular behavior of the pitch angles of the two brighter inner spirals could be hinting at a second companion hidden in the inner disk regions not probed by our datasets, and possibly in the 7\,au cavity inferred by \cite{2008ApJ...684.1323P}. 

%%%%%%%%%%%%%%%%%%%%%%%%%%%%%%%%%%%%%%%%%%%%%%%%%%
% Acknowledgement
%%%%%%%%%%%%%%%%%%%%%%%%%%%%%%%%%%%%%%%%%%%%%%%%%%

\section*{Acknowledgements}

SPHERE is an instrument designed and built by a consortium
consisting of IPAG (Grenoble, France), MPIA (Heidelberg, Germany),
LAM (Marseille, France), LESIA (Paris, France), Laboratoire Lagrange
(Nice, France), INAF - Osservatorio di Padova (Italy), Observatoire de
Genève (Switzerland), ETH Zurich (Switzerland), NOVA (Netherlands), ONERA
(France), and ASTRON (The Netherlands) in collaboration with ESO.
SPHERE was funded by ESO, with additional contributions from CNRS
(France), MPIA (Germany), INAF (Italy), FINES (Switzerland), and NOVA
(The Netherlands). SPHERE also received funding from the European Commission
Sixth and Seventh Framework Programmes as part of the Optical Infrared
Coordination Network for Astronomy (OPTICON) under grant number RII3-Ct2004-001566
for FP6 (2004-2008), grant number 226604 for FP7 (2009-2012),
and grant number 312430 for FP7 (2013-2016). G.M-A. and C.G. acknowledge funding from the Netherlands Organisation for Scientific Research (NWO) TOP-1 grant as part
of the research program “Herbig Ae/Be stars, Rosetta stones for understanding
the formation of planetary systems”, project number 614.001.552.
FMe, MV, and MB acknowledge funding from ANR of France under contract number ANR-16-CE31-0013.
The research of AJB leading to these results has received funding from the European Research Council under ERC Starting Grant agreement 678194 (FALCONER).
P.P. acknowledges support provided by the Alexander von Humboldt Foundation in the framework of the Sofja Kovalevskaja Award endowed by the Federal Ministry of Education and Research.
D.H. is supported by European Union A-ERC grant 291141 CHEMPLAN, NWO and by a KNAW professor prize awarded to E. van Dishoeck.
T.H. acknowledges support from the European Research Council under the Horizon 2020 Framework Program via the ERC Advanced Grant Origins 83 24 28.
M.T. has been supported by the UK Science and Technology research Council (STFC), and by the European Union’s Horizon 2020 research and innovation programme under the Marie Sklodowska-Curie grant agreement No. 823823 (RISE DUSTBUSTERS project).

\begin{appendix}

\section{Keck L-band detection limits}
\label{keck-appendix}

\begin{figure}[ht]
\center
%\begin{tabular}{ccc}
\includegraphics[width=0.48\textwidth]{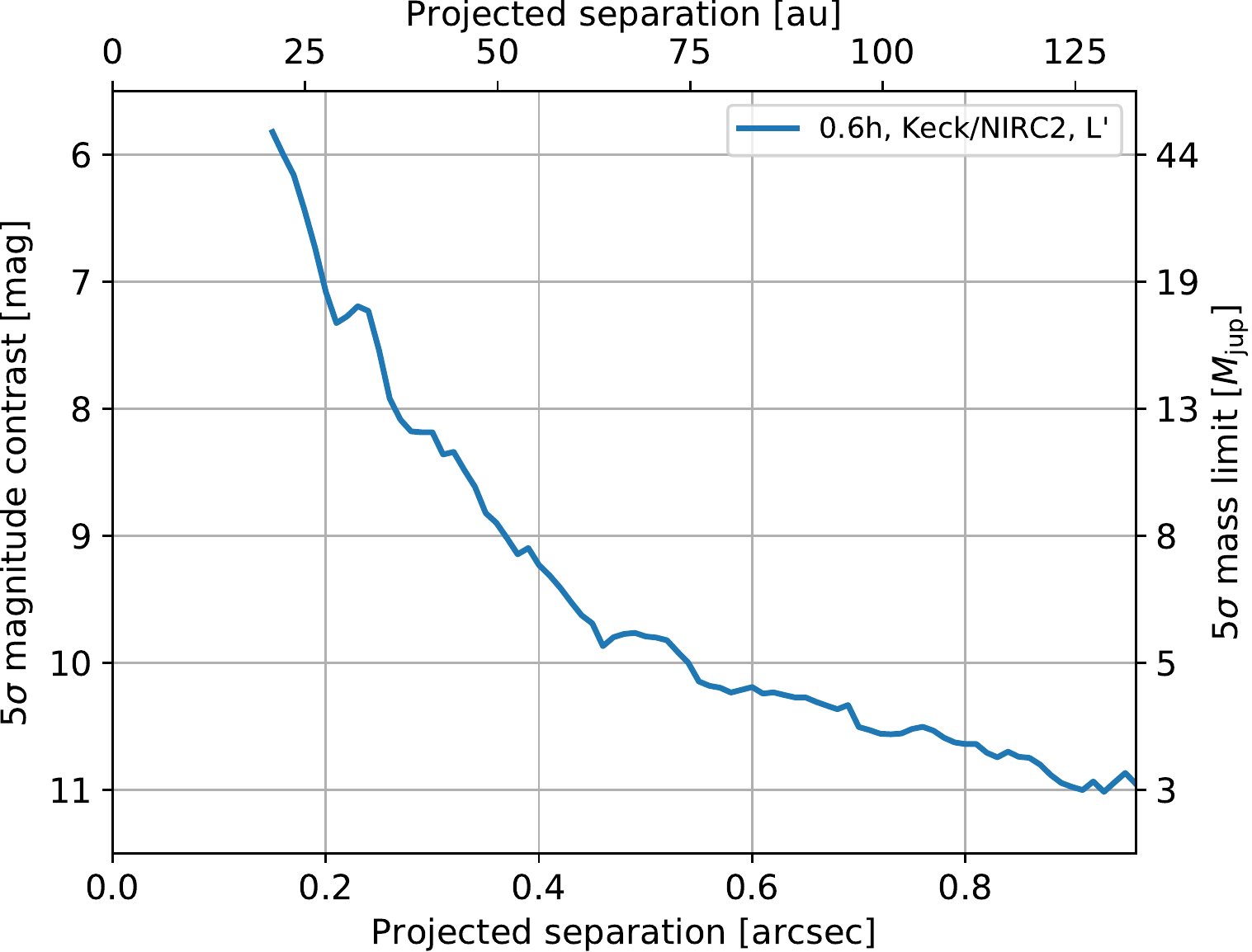} 
\caption{Detection limits derived from Keck/NIRC2 L-band observations of SR\,21.} 
\label{fig:Appendix:L-band}
\end{figure}

SR\,21 was observed (PI: N. van der Marel) with the NIRC2 camera that is mounted at the Naysmith platform of the Keck II telescope at the W.M. Keck Observatory. The observations were carried out on the night of April 23, 2016, in L’ band and no coronagraph was applied. A sequence of 48 unsaturated images with exposure times of 0.2\,s and 100 co-added frames was acquired before the science observations consisting of 120 frames with exposure times of 0.3s and 60 co-adds allowing for a field rotation of 18\fdg19. The data were reduced with PynPoint (v. 0.8.1; \citealt{2012MNRAS.427..948A, 2019A&A...621A..59S}) following the description for NACO data presented in \cite{2020MNRAS.492..431B}. This included dark, flat calibration, bad pixel and background subtraction based on principal component analysis (PCA; \citealt{2018A&A...611A..23H}). The PSFs were centered by a Gaussian fit and we applied ADI+PCA for speckle removal. The contrast as a function of separation was evaluated with the \texttt{ContrastCurveModule} of PynPoint. We scaled the average unsaturated PSF image for the difference in exposure time and injected it in the centered science frames. The contrast was estimated at six azimuthal directions at angular separations increasing from 150mas to 960mas in steps of 10\,mas. We optimized the number of subtracted principal components as a function of separation. The final curve as presented in Figure~\ref{fig:Appendix:L-band} was obtained by fitting five components for separations smaller than 250mas and ten components for larger separations. We further converted the magnitude contrast to a threshold of detectable masses, by comparison to AMES Cond models evaluated at the system age of 10\,Myr. For this conversion we assumed a stellar L’ magnitude of 6.8\,mag based on WISE measurements.

\end{appendix}

\bibliographystyle{aa}
\bibliography{sr21_bib} 

\begin{thebibliography}{53}
\expandafter\ifx\csname natexlab\endcsname\relax\def\natexlab#1{#1}\fi

\bibitem[{{Akiyama} {et~al.}(2016){Akiyama}, {Hashimoto}, {Liu}, {Li},
  {Bonnefoy}, {Dong}, {Hasegawa}, {Henning}, {Sitko}, {Janson}, {Feldt},
  {Wisniewski}, {Kudo}, {Kusakabe}, {Tsukagoshi}, {Momose}, {Muto}, {Taki},
  {Kuzuhara}, {Satoshi}, {Takami}, {Ohashi}, {Grady}, {Kwon}, {Thalmann},
  {Abe}, {Brandner}, {Brand t}, {Carson}, {Egner}, {Goto}, {Guyon}, {Hayano},
  {Hayashi}, {Hayashi}, {Hodapp}, {Ishii}, {Iye}, {Knapp}, {Kand ori},
  {Matsuo}, {Mcelwain}, {Miyama}, {Morino}, {Moro-Martin}, {Nishimura}, {Pyo},
  {Serabyn}, {Suenaga}, {Suto}, {Suzuki}, {Takahashi}, {Takato}, {Terada},
  {Tomono}, {Turner}, {Watanabe}, {Yamada}, {Takami}, {Usuda}, \&
  {Tamura}}]{2016AJ....152..222A}
{Akiyama}, E., {Hashimoto}, J., {Liu}, H.~B., {et~al.} 2016, \aj, 152, 222

\bibitem[{{Amara} \& {Quanz}(2012)}]{2012MNRAS.427..948A}
{Amara}, A. \& {Quanz}, S.~P. 2012, \mnras, 427, 948

\bibitem[{{Andrews} {et~al.}(2018){Andrews}, {Huang}, {P{\'e}rez}, {Isella},
  {Dullemond}, {Kurtovic}, {Guzm{\'a}n}, {Carpenter}, {Wilner}, {Zhang}, {Zhu},
  {Birnstiel}, {Bai}, {Benisty}, {Hughes}, {{\"O}berg}, \&
  {Ricci}}]{2018ApJ...869L..41A}
{Andrews}, S.~M., {Huang}, J., {P{\'e}rez}, L.~M., {et~al.} 2018, \apjl, 869,
  L41

\bibitem[{{Andrews} {et~al.}(2011){Andrews}, {Wilner}, {Espaillat}, {Hughes},
  {Dullemond}, {McClure}, {Qi}, \& {Brown}}]{2011ApJ...732...42A}
{Andrews}, S.~M., {Wilner}, D.~J., {Espaillat}, C., {et~al.} 2011, \apj, 732,
  42

\bibitem[{{Ataiee} {et~al.}(2013){Ataiee}, {Pinilla}, {Zsom}, {Dullemond},
  {Dominik}, \& {Ghanbari}}]{2013A&A...553L...3A}
{Ataiee}, S., {Pinilla}, P., {Zsom}, A., {et~al.} 2013, \aap, 553, L3

\bibitem[{{Avenhaus} {et~al.}(2018){Avenhaus}, {Quanz}, {Garufi}, {Perez},
  {Casassus}, {Pinte}, {Bertrang}, {Caceres}, {Benisty}, \&
  {Dominik}}]{2018ApJ...863...44A}
{Avenhaus}, H., {Quanz}, S.~P., {Garufi}, A., {et~al.} 2018, \apj, 863, 44

\bibitem[{{Barsony} {et~al.}(2003){Barsony}, {Koresko}, \&
  {Matthews}}]{2003ApJ...591.1064B}
{Barsony}, M., {Koresko}, C., \& {Matthews}, K. 2003, \apj, 591, 1064

\bibitem[{{Benisty} {et~al.}(2015){Benisty}, {Juhasz}, {Boccaletti},
  {Avenhaus}, {Milli}, {Thalmann}, {Dominik}, {Pinilla}, {Buenzli}, {Pohl},
  {Beuzit}, {Birnstiel}, {de Boer}, {Bonnefoy}, {Chauvin}, {Christiaens},
  {Garufi}, {Grady}, {Henning}, {Huelamo}, {Isella}, {Langlois}, {M{\'e}nard},
  {Mouillet}, {Olofsson}, {Pantin}, {Pinte}, \& {Pueyo}}]{2015A&A...578L...6B}
{Benisty}, M., {Juhasz}, A., {Boccaletti}, A., {et~al.} 2015, \aap, 578, L6

\bibitem[{{Beuzit} {et~al.}(2019){Beuzit}, {Vigan}, {Mouillet}, {Dohlen},
  {Gratton}, {Boccaletti}, {Sauvage}, {Schmid}, {Langlois}, {Petit},
  {Baruffolo}, {Feldt}, {Milli}, {Wahhaj}, {Abe}, {Anselmi}, {Antichi},
  {Barette}, {Baudrand}, {Baudoz}, {Bazzon}, {Bernardi}, {Blanchard}, {Brast},
  {Bruno}, {Buey}, {Carbillet}, {Carle}, {Cascone}, {Chapron}, {Charton},
  {Chauvin}, {Claudi}, {Costille}, {De Caprio}, {de Boer}, {Delboulb{\'e}},
  {Desidera}, {Dominik}, {Downing}, {Dupuis}, {Fabron}, {Fantinel}, {Farisato},
  {Feautrier}, {Fedrigo}, {Fusco}, {Gigan}, {Ginski}, {Girard}, {Giro},
  {Gisler}, {Gluck}, {Gry}, {Henning}, {Hubin}, {Hugot}, {Incorvaia}, {Jaquet},
  {Kasper}, {Lagadec}, {Lagrange}, {Le Coroller}, {Le Mignant}, {Le Ruyet},
  {Lessio}, {Lizon}, {Llored}, {Lundin}, {Madec}, {Magnard}, {Marteaud},
  {Martinez}, {Maurel}, {M{\'e}nard}, {Mesa}, {M{\"o}ller-Nilsson}, {Moulin},
  {Moutou}, {Orign{\'e}}, {Parisot}, {Pavlov}, {Perret}, {Pragt}, {Puget},
  {Rabou}, {Ramos}, {Reess}, {Rigal}, {Rochat}, {Roelfsema}, {Rousset}, {Roux},
  {Saisse}, {Salasnich}, {Santambrogio}, {Scuderi}, {Segransan}, {Sevin},
  {Siebenmorgen}, {Soenke}, {Stadler}, {Suarez}, {Tiph{\`e}ne}, {Turatto},
  {Udry}, {Vakili}, {Waters}, {Weber}, {Wildi}, {Zins}, \&
  {Zurlo}}]{2019A&A...631A.155B}
{Beuzit}, J.~L., {Vigan}, A., {Mouillet}, D., {et~al.} 2019, \aap, 631, A155

\bibitem[{{Bohn} {et~al.}(2020){Bohn}, {Kenworthy}, {Ginski}, {Manara},
  {Pecaut}, {de Boer}, {Keller}, {Mamajek}, {Meshkat}, {Reggiani}, {Todorov},
  \& {Snik}}]{2020MNRAS.492..431B}
{Bohn}, A.~J., {Kenworthy}, M.~A., {Ginski}, C., {et~al.} 2020, \mnras, 492,
  431

\bibitem[{{Brown} {et~al.}(2007){Brown}, {Blake}, {Dullemond}, {Mer{\'{\i}}n},
  {Augereau}, {Boogert}, {Evans}, {Geers}, {Lahuis}, {Kessler-Silacci},
  {Pontoppidan}, \& {van Dishoeck}}]{2007ApJ...664L.107B}
{Brown}, J.~M., {Blake}, G.~A., {Dullemond}, C.~P., {et~al.} 2007, \apjl, 664,
  L107

\bibitem[{{Carbillet} {et~al.}(2011){Carbillet}, {Bendjoya}, {Abe}, {Guerri},
  {Boccaletti}, {Daban}, {Dohlen}, {Ferrari}, {Robbe-Dubois}, {Douet}, \&
  {Vakili}}]{2011ExA....30...39C}
{Carbillet}, M., {Bendjoya}, P., {Abe}, L., {et~al.} 2011, Experimental
  Astronomy, 30, 39

\bibitem[{{Cazzoletti} {et~al.}(2018){Cazzoletti}, {van Dishoeck}, {Pinilla},
  {Tazzari}, {Facchini}, {van der Marel}, {Benisty}, {Garufi}, \&
  {P{\'e}rez}}]{2018A&A...619A.161C}
{Cazzoletti}, P., {van Dishoeck}, E.~F., {Pinilla}, P., {et~al.} 2018, \aap,
  619, A161

\bibitem[{{de Boer} {et~al.}(2020){de Boer}, {Langlois}, {van Holstein},
  {Girard}, {Mouillet}, {Vigan}, {Dohlen}, {Snik}, {Keller}, {Ginski}, {Stam},
  {Milli}, {Wahhaj}, {Kasper}, {Schmid}, {Rabou}, {Gluck}, {Hugot}, {Perret},
  {Martinez}, {Weber}, {Pragt}, {Sauvage}, {Boccaletti}, {Le Coroller},
  {Dominik}, {Henning}, {Lagadec}, {M{\'e}nard}, {Turatto}, {Udry}, {Chauvin},
  {Feldt}, \& {Beuzit}}]{deBoer2020}
{de Boer}, J., {Langlois}, M., {van Holstein}, R.~G., {et~al.} 2020, \aap, 633,
  A63

\bibitem[{{de Juan Ovelar} {et~al.}(2013){de Juan Ovelar}, {Min}, {Dominik},
  {Thalmann}, {Pinilla}, {Benisty}, \& {Birnstiel}}]{2013A&A...560A.111D}
{de Juan Ovelar}, M., {Min}, M., {Dominik}, C., {et~al.} 2013, \aap, 560, A111

\bibitem[{{Dohlen} {et~al.}(2008){Dohlen}, {Langlois}, {Saisse}, {Hill},
  {Origne}, {Jacquet}, {Fabron}, {Blanc}, {Llored}, {Carle}, {Moutou}, {Vigan},
  {Boccaletti}, {Carbillet}, {Mouillet}, \& {Beuzit}}]{2008SPIE.7014E..3LD}
{Dohlen}, K., {Langlois}, M., {Saisse}, M., {et~al.} 2008, in \procspie, Vol.
  7014, Ground-based and Airborne Instrumentation for Astronomy II, 70143L

\bibitem[{{Dong} {et~al.}(2015){Dong}, {Zhu}, {Rafikov}, \&
  {Stone}}]{2015ApJ...809L...5D}
{Dong}, R., {Zhu}, Z., {Rafikov}, R.~R., \& {Stone}, J.~M. 2015, \apjl, 809, L5

\bibitem[{{Dullemond} {et~al.}(2018){Dullemond}, {Birnstiel}, {Huang},
  {Kurtovic}, {Andrews}, {Guzm{\'a}n}, {P{\'e}rez}, {Isella}, {Zhu}, {Benisty},
  {Wilner}, {Bai}, {Carpenter}, {Zhang}, \& {Ricci}}]{2018ApJ...869L..46D}
{Dullemond}, C.~P., {Birnstiel}, T., {Huang}, J., {et~al.} 2018, \apjl, 869,
  L46

\bibitem[{{Eisner} {et~al.}(2009){Eisner}, {Monnier}, {Tuthill}, \&
  {Lacour}}]{2009ApJ...698L.169E}
{Eisner}, J.~A., {Monnier}, J.~D., {Tuthill}, P., \& {Lacour}, S. 2009, \apjl,
  698, L169

\bibitem[{{Follette} {et~al.}(2013){Follette}, {Tamura}, {Hashimoto},
  {Whitney}, {Grady}, {Close}, {Andrews}, {Kwon}, {Wisniewski}, {Brandt},
  {Mayama}, {Kandori}, {Dong}, {Abe}, {Brandner}, {Carson}, {Currie}, {Egner},
  {Feldt}, {Goto}, {Guyon}, {Hayano}, {Hayashi}, {Hayashi}, {Henning},
  {Hodapp}, {Ishii}, {Iye}, {Janson}, {Knapp}, {Kudo}, {Kusakabe}, {Kuzuhara},
  {McElwain}, {Matsuo}, {Miyama}, {Morino}, {Moro-Martin}, {Nishimura}, {Pyo},
  {Serabyn}, {Suto}, {Suzuki}, {Takami}, {Takato}, {Terada}, {Thalmann},
  {Tomono}, {Turner}, {Watanabe}, {Yamada}, {Takami}, \&
  {Usuda}}]{2013ApJ...767...10F}
{Follette}, K.~B., {Tamura}, M., {Hashimoto}, J., {et~al.} 2013, \apj, 767, 10

\bibitem[{{Gaia Collaboration}(2018)}]{2018yCat.1345....0G}
{Gaia Collaboration}. 2018, VizieR Online Data Catalog, I/345

\bibitem[{{Garufi} {et~al.}(2013){Garufi}, {Quanz}, {Avenhaus}, {Buenzli},
  {Dominik}, {Meru}, {Meyer}, {Pinilla}, {Schmid}, \&
  {Wolf}}]{2013A&A...560A.105G}
{Garufi}, A., {Quanz}, S.~P., {Avenhaus}, H., {et~al.} 2013, \aap, 560, A105

\bibitem[{{Ginski} {et~al.}(2016){Ginski}, {Stolker}, {Pinilla}, {Dominik},
  {Boccaletti}, {de Boer}, {Benisty}, {Biller}, {Feldt}, {Garufi}, {Keller},
  {Kenworthy}, {Maire}, {M{\'e}nard}, {Mesa}, {Milli}, {Min}, {Pinte}, {Quanz},
  {van Boekel}, {Bonnefoy}, {Chauvin}, {Desidera}, {Gratton}, {Girard},
  {Keppler}, {Kopytova}, {Lagrange}, {Langlois}, {Rouan}, \&
  {Vigan}}]{2016A&A...595A.112G}
{Ginski}, C., {Stolker}, T., {Pinilla}, P., {et~al.} 2016, \aap, 595, A112

\bibitem[{{Grady} {et~al.}(2013){Grady}, {Muto}, {Hashimoto}, {Fukagawa},
  {Currie}, {Biller}, {Thalmann}, {Sitko}, {Russell}, {Wisniewski}, {Dong},
  {Kwon}, {Sai}, {Hornbeck}, {Schneider}, {Hines}, {Moro Mart{\'\i}n}, {Feldt},
  {Henning}, {Pott}, {Bonnefoy}, {Bouwman}, {Lacour}, {Mueller}, {Juh{\'a}sz},
  {Crida}, {Chauvin}, {Andrews}, {Wilner}, {Kraus}, {Dahm}, {Robitaille},
  {Jang-Condell}, {Abe}, {Akiyama}, {Brandner}, {Brandt}, {Carson}, {Egner},
  {Follette}, {Goto}, {Guyon}, {Hayano}, {Hayashi}, {Hayashi}, {Hodapp},
  {Ishii}, {Iye}, {Janson}, {Kandori}, {Knapp}, {Kudo}, {Kusakabe}, {Kuzuhara},
  {Mayama}, {McElwain}, {Matsuo}, {Miyama}, {Morino}, {Nishimura}, {Pyo},
  {Serabyn}, {Suto}, {Suzuki}, {Takami}, {Takato}, {Terada}, {Tomono},
  {Turner}, {Watanabe}, {Yamada}, {Takami}, {Usuda}, \&
  {Tamura}}]{2013ApJ...762...48G}
{Grady}, C.~A., {Muto}, T., {Hashimoto}, J., {et~al.} 2013, \apj, 762, 48

\bibitem[{{Gratton} {et~al.}(2019){Gratton}, {Ligi}, {Sissa}, {Desidera},
  {Mesa}, {Bonnefoy}, {Chauvin}, {Cheetham}, {Feldt}, {Lagrange}, {Langlois},
  {Meyer}, {Vigan}, {Boccaletti}, {Janson}, {Lazzoni}, {Zurlo}, {De Boer},
  {Henning}, {D'Orazi}, {Gluck}, {Madec}, {Jaquet}, {Baudoz}, {Fantinel},
  {Pavlov}, \& {Wildi}}]{2019A&A...623A.140G}
{Gratton}, R., {Ligi}, R., {Sissa}, E., {et~al.} 2019, \aap, 623, A140

\bibitem[{{Haffert} {et~al.}(2019){Haffert}, {Bohn}, {de Boer}, {Snellen},
  {Brinchmann}, {Girard}, {Keller}, \& {Bacon}}]{2019NatAs...3..749H}
{Haffert}, S.~Y., {Bohn}, A.~J., {de Boer}, J., {et~al.} 2019, Nature
  Astronomy, 3, 749

\bibitem[{{Herczeg} \& {Hillenbrand}(2014)}]{2014ApJ...786...97H}
{Herczeg}, G.~J. \& {Hillenbrand}, L.~A. 2014, \apj, 786, 97

\bibitem[{{Huang} {et~al.}(2018){Huang}, {Andrews}, {Dullemond}, {Isella},
  {P{\'e}rez}, {Guzm{\'a}n}, {{\"O}berg}, {Zhu}, {Zhang}, {Bai}, {Benisty},
  {Birnstiel}, {Carpenter}, {Hughes}, {Ricci}, {Weaver}, \&
  {Wilner}}]{2018ApJ...869L..42H}
{Huang}, J., {Andrews}, S.~M., {Dullemond}, C.~P., {et~al.} 2018, \apjl, 869,
  L42

\bibitem[{{Hunziker} {et~al.}(2018){Hunziker}, {Quanz}, {Amara}, \&
  {Meyer}}]{2018A&A...611A..23H}
{Hunziker}, S., {Quanz}, S.~P., {Amara}, A., \& {Meyer}, M.~R. 2018, \aap, 611,
  A23

\bibitem[{{Keppler} {et~al.}(2018){Keppler}, {Benisty}, {M{\"u}ller},
  {Henning}, {van Boekel}, {Cantalloube}, {Ginski}, {van Holstein}, {Maire},
  {Pohl}, {Samland}, {Avenhaus}, {Baudino}, {Boccaletti}, {de Boer},
  {Bonnefoy}, {Chauvin}, {Desidera}, {Langlois}, {Lazzoni}, {Marleau},
  {Mordasini}, {Pawellek}, {Stolker}, {Vigan}, {Zurlo}, {Birnstiel},
  {Brandner}, {Feldt}, {Flock}, {Girard}, {Gratton}, {Hagelberg}, {Isella},
  {Janson}, {Juhasz}, {Kemmer}, {Kral}, {Lagrange}, {Launhardt}, {Matter},
  {M{\'e}nard}, {Milli}, {Molli{\`e}re}, {Olofsson}, {P{\'e}rez}, {Pinilla},
  {Pinte}, {Quanz}, {Schmidt}, {Udry}, {Wahhaj}, {Williams}, {Buenzli},
  {Cudel}, {Dominik}, {Galicher}, {Kasper}, {Lannier}, {Mesa}, {Mouillet},
  {Peretti}, {Perrot}, {Salter}, {Sissa}, {Wildi}, {Abe}, {Antichi},
  {Augereau}, {Baruffolo}, {Baudoz}, {Bazzon}, {Beuzit}, {Blanchard}, {Brems},
  {Buey}, {De Caprio}, {Carbillet}, {Carle}, {Cascone}, {Cheetham}, {Claudi},
  {Costille}, {Delboulb{\'e}}, {Dohlen}, {Fantinel}, {Feautrier}, {Fusco},
  {Giro}, {Gluck}, {Gry}, {Hubin}, {Hugot}, {Jaquet}, {Le Mignant}, {Llored},
  {Madec}, {Magnard}, {Martinez}, {Maurel}, {Meyer}, {M{\"o}ller-Nilsson},
  {Moulin}, {Mugnier}, {Orign{\'e}}, {Pavlov}, {Perret}, {Petit}, {Pragt},
  {Puget}, {Rabou}, {Ramos}, {Rigal}, {Rochat}, {Roelfsema}, {Rousset}, {Roux},
  {Salasnich}, {Sauvage}, {Sevin}, {Soenke}, {Stadler}, {Suarez}, {Turatto}, \&
  {Weber}}]{2018A&A...617A..44K}
{Keppler}, M., {Benisty}, M., {M{\"u}ller}, A., {et~al.} 2018, \aap, 617, A44

\bibitem[{{Kraus} {et~al.}(2017){Kraus}, {Kreplin}, {Fukugawa}, {Muto},
  {Sitko}, {Young}, {Bate}, {Grady}, {Harries}, {Monnier}, {Willson}, \&
  {Wisniewski}}]{2017ApJ...848L..11K}
{Kraus}, S., {Kreplin}, A., {Fukugawa}, M., {et~al.} 2017, \apjl, 848, L11

\bibitem[{{Langlois} {et~al.}(2014){Langlois}, {Dohlen}, {Vigan}, {Zurlo},
  {Moutou}, {Schmid}, {Mili}, {Beuzit}, {Boccaletti}, {Carle}, {Costille},
  {Dorn}, {Gluck}, {Hubin}, {Feldt}, {Kasper}, {Lizon}, {Madec}, {Le Mignant},
  {Mouillet}, {Puget}, {Sauvage}, \& {Wildi}}]{2014SPIE.9147E..1RL}
{Langlois}, M., {Dohlen}, K., {Vigan}, A., {et~al.} 2014, in \procspie, Vol.
  9147, Ground-based and Airborne Instrumentation for Astronomy V, 91471R

\bibitem[{{Manara} {et~al.}(2015){Manara}, {Testi}, {Natta}, \&
  {Alcal{\'a}}}]{2015A&A...579A..66M}
{Manara}, C.~F., {Testi}, L., {Natta}, A., \& {Alcal{\'a}}, J.~M. 2015, \aap,
  579, A66

\bibitem[{{Martinez} {et~al.}(2009){Martinez}, {Dorrer}, {Aller Carpentier},
  {Kasper}, {Boccaletti}, {Dohlen}, \& {Yaitskova}}]{2009A&A...495..363M}
{Martinez}, P., {Dorrer}, C., {Aller Carpentier}, E., {et~al.} 2009, \aap, 495,
  363

\bibitem[{{McMullin} {et~al.}(2007){McMullin}, {Waters}, {Schiebel}, {Young},
  \& {Golap}}]{2007ASPC..376..127M}
{McMullin}, J.~P., {Waters}, B., {Schiebel}, D., {Young}, W., \& {Golap}, K.
  2007, Astronomical Society of the Pacific Conference Series, Vol. 376, {CASA
  Architecture and Applications}, ed. R.~A. {Shaw}, F.~{Hill}, \& D.~J. {Bell},
  127

\bibitem[{{M{\"u}ller} {et~al.}(2018){M{\"u}ller}, {Keppler}, {Henning},
  {Samland}, {Chauvin}, {Beust}, {Maire}, {Molaverdikhani}, {van Boekel},
  {Benisty}, {Boccaletti}, {Bonnefoy}, {Cantalloube}, {Charnay}, {Baudino},
  {Gennaro}, {Long}, {Cheetham}, {Desidera}, {Feldt}, {Fusco}, {Girard},
  {Gratton}, {Hagelberg}, {Janson}, {Lagrange}, {Langlois}, {Lazzoni}, {Ligi},
  {M{\'e}nard}, {Mesa}, {Meyer}, {Molli{\`e}re}, {Mordasini}, {Moulin},
  {Pavlov}, {Pawellek}, {Quanz}, {Ramos}, {Rouan}, {Sissa}, {Stadler}, {Vigan},
  {Wahhaj}, {Weber}, \& {Zurlo}}]{2018A&A...617L...2M}
{M{\"u}ller}, A., {Keppler}, M., {Henning}, T., {et~al.} 2018, \aap, 617, L2

\bibitem[{{Muto} {et~al.}(2012){Muto}, {Grady}, {Hashimoto}, {Fukagawa},
  {Hornbeck}, {Sitko}, {Russell}, {Werren}, {Cur{\'e}}, {Currie}, {Ohashi},
  {Okamoto}, {Momose}, {Honda}, {Inutsuka}, {Takeuchi}, {Dong}, {Abe},
  {Brandner}, {Brandt}, {Carson}, {Egner}, {Feldt}, {Fukue}, {Goto}, {Guyon},
  {Hayano}, {Hayashi}, {Hayashi}, {Henning}, {Hodapp}, {Ishii}, {Iye},
  {Janson}, {Kandori}, {Knapp}, {Kudo}, {Kusakabe}, {Kuzuhara}, {Matsuo},
  {Mayama}, {McElwain}, {Miyama}, {Morino}, {Moro-Martin}, {Nishimura}, {Pyo},
  {Serabyn}, {Suto}, {Suzuki}, {Takami}, {Takato}, {Terada}, {Thalmann},
  {Tomono}, {Turner}, {Watanabe}, {Wisniewski}, {Yamada}, {Takami}, {Usuda}, \&
  {Tamura}}]{2012ApJ...748L..22M}
{Muto}, T., {Grady}, C.~A., {Hashimoto}, J., {et~al.} 2012, \apjl, 748, L22

\bibitem[{{Nelson} {et~al.}(2000){Nelson}, {Papaloizou}, {Masset}, \&
  {Kley}}]{2000MNRAS.318...18N}
{Nelson}, R.~P., {Papaloizou}, J. C.~B., {Masset}, F., \& {Kley}, W. 2000,
  \mnras, 318, 18

\bibitem[{{P{\'e}rez} {et~al.}(2014){P{\'e}rez}, {Isella}, {Carpenter}, \&
  {Chandler}}]{2014ApJ...783L..13P}
{P{\'e}rez}, L.~M., {Isella}, A., {Carpenter}, J.~M., \& {Chandler}, C.~J.
  2014, \apjl, 783, L13

\bibitem[{{Pinilla} {et~al.}(2015{\natexlab{a}}){Pinilla}, {de Juan Ovelar},
  {Ataiee}, {Benisty}, {Birnstiel}, {van Dishoeck}, \& {Min}}]{pinilla2015}
{Pinilla}, P., {de Juan Ovelar}, M., {Ataiee}, S., {et~al.} 2015{\natexlab{a}},
  \aap, 573, A9

\bibitem[{{Pinilla} {et~al.}(2015{\natexlab{b}}){Pinilla}, {van der Marel},
  {P{\'e}rez}, {van Dishoeck}, {Andrews}, {Birnstiel}, {Herczeg},
  {Pontoppidan}, \& {van Kempen}}]{2015A&A...584A..16P}
{Pinilla}, P., {van der Marel}, N., {P{\'e}rez}, L.~M., {et~al.}
  2015{\natexlab{b}}, \aap, 584, A16

\bibitem[{{Pontoppidan} {et~al.}(2008){Pontoppidan}, {Blake}, {van Dishoeck},
  {Smette}, {Ireland}, \& {Brown}}]{2008ApJ...684.1323P}
{Pontoppidan}, K.~M., {Blake}, G.~A., {van Dishoeck}, E.~F., {et~al.} 2008,
  \apj, 684, 1323

\bibitem[{{Rosotti} {et~al.}(2019){Rosotti}, {Benisty}, {Juh{\'a}sz}, {Teague},
  {Clarke}, {Dominik}, {Dullemond}, {Klaassen}, {Matra}, \&
  {Stolker}}]{2019MNRAS.tmp.2689R}
{Rosotti}, G.~P., {Benisty}, M., {Juh{\'a}sz}, A., {et~al.} 2019, \mnras, 2689

\bibitem[{{Rosotti} {et~al.}(2016){Rosotti}, {Juhasz}, {Booth}, \&
  {Clarke}}]{2016MNRAS.459.2790R}
{Rosotti}, G.~P., {Juhasz}, A., {Booth}, R.~A., \& {Clarke}, C.~J. 2016,
  \mnras, 459, 2790

\bibitem[{{Sallum} {et~al.}(2019){Sallum}, {Skemer}, {Eisner}, {van der Marel},
  {Sheehan}, {Close}, {Ireland}, {Males}, {Morzinski}, {Bailey}, {Briguglio},
  \& {Puglisi}}]{2019arXiv190807427S}
{Sallum}, S., {Skemer}, A., {Eisner}, J., {et~al.} 2019, arXiv e-prints,
  arXiv:1908.07427

\bibitem[{{S{\'a}ndor} {et~al.}(2007){S{\'a}ndor}, {Kley}, \&
  {Klagyivik}}]{2007A&A...472..981S}
{S{\'a}ndor}, Z., {Kley}, W., \& {Klagyivik}, P. 2007, \aap, 472, 981

\bibitem[{{Schmid} {et~al.}(2006){Schmid}, {Joos}, \&
  {Tschan}}]{2006A&A...452..657S}
{Schmid}, H.~M., {Joos}, F., \& {Tschan}, D. 2006, \aap, 452, 657

\bibitem[{{Stolker} {et~al.}(2019){Stolker}, {Bonse}, {Quanz}, {Amara},
  {Cugno}, {Bohn}, \& {Boehle}}]{2019A&A...621A..59S}
{Stolker}, T., {Bonse}, M.~J., {Quanz}, S.~P., {et~al.} 2019, \aap, 621, A59

\bibitem[{{Stolker} {et~al.}(2016){Stolker}, {Dominik}, {Avenhaus}, {Min}, {de
  Boer}, {Ginski}, {Schmid}, {Juhasz}, {Bazzon}, {Waters}, {Garufi},
  {Augereau}, {Benisty}, {Boccaletti}, {Henning}, {Langlois}, {Maire},
  {M{\'e}nard}, {Meyer}, {Pinte}, {Quanz}, {Thalmann}, {Beuzit}, {Carbillet},
  {Costille}, {Dohlen}, {Feldt}, {Gisler}, {Mouillet}, {Pavlov}, {Perret},
  {Petit}, {Pragt}, {Rochat}, {Roelfsema}, {Salasnich}, {Soenke}, \&
  {Wildi}}]{2016A&A...595A.113S}
{Stolker}, T., {Dominik}, C., {Avenhaus}, H., {et~al.} 2016, \aap, 595, A113

\bibitem[{{Uyama} {et~al.}(2018){Uyama}, {Hashimoto}, {Muto}, {Akiyama},
  {Dong}, {de Leon}, {Sakon}, {Kudo}, {Kusakabe}, {Kuzuhara}, {Bonnefoy},
  {Abe}, {Brand ner}, {Brandt}, {Carson}, {Currie}, {Egner}, {Feldt}, {Fung},
  {Goto}, {Grady}, {Guyon}, {Hayano}, {Hayashi}, {Hayashi}, {Henning},
  {Hodapp}, {Ishii}, {Iye}, {Janson}, {Kand ori}, {Knapp}, {Kwon}, {Matsuo},
  {Mayama}, {Mcelwain}, {Miyama}, {Morino}, {Moro-Martin}, {Nishimura}, {Pyo},
  {Serabyn}, {Sitko}, {Suenaga}, {Suto}, {Suzuki}, {Takahashi}, {Takami},
  {Takato}, {Terada}, {Thalmann}, {Turner}, {Watanabe}, {Wisniewski}, {Yamada},
  {Yang}, {Takami}, {Usuda}, \& {Tamura}}]{2018AJ....156...63U}
{Uyama}, T., {Hashimoto}, J., {Muto}, T., {et~al.} 2018, \aj, 156, 63

\bibitem[{{van der Marel} {et~al.}(2016){van der Marel}, {van Dishoeck},
  {Bruderer}, {Andrews}, {Pontoppidan}, {Herczeg}, {van Kempen}, \&
  {Miotello}}]{2016A&A...585A..58V}
{van der Marel}, N., {van Dishoeck}, E.~F., {Bruderer}, S., {et~al.} 2016,
  \aap, 585, A58

\bibitem[{{van Holstein} {et~al.}(2020){van Holstein}, {Girard}, {de Boer},
  {Snik}, {Milli}, {Stam}, {Ginski}, {Mouillet}, {Wahhaj}, {Schmid}, {Keller},
  {Langlois}, {Dohlen}, {Vigan}, {Pohl}, {Carbillet}, {Fantinel}, {Maurel},
  {Orign{\'e}}, {Petit}, {Ramos}, {Rigal}, {Sevin}, {Boccaletti}, {Le
  Coroller}, {Dominik}, {Henning}, {Lagadec}, {M{\'e}nard}, {Turatto}, {Udry},
  {Chauvin}, {Feldt}, \& {Beuzit}}]{vanHolstein2020}
{van Holstein}, R.~G., {Girard}, J.~H., {de Boer}, J., {et~al.} 2020, \aap,
  633, A64

\bibitem[{{van Holstein} {et~al.}(2017){van Holstein}, {Snik}, {Girard}, {de
  Boer}, {Ginski}, {Keller}, {Stam}, {Beuzit}, {Mouillet}, {Kasper},
  {Langlois}, {Zurlo}, {de Kok}, \& {Vigan}}]{2017SPIE10400E..15V}
{van Holstein}, R.~G., {Snik}, F., {Girard}, J.~H., {et~al.} 2017, in Society
  of Photo-Optical Instrumentation Engineers (SPIE) Conference Series, Vol.
  10400, Society of Photo-Optical Instrumentation Engineers (SPIE) Conference
  Series, 1040015

\end{thebibliography}

\end{document}